\documentclass[a4paper]{article}
\usepackage{Odyssey2020}
\usepackage{epsfig,amssymb,amsmath}
\usepackage{graphicx, epstopdf, dblfloatfix, subfig}
\usepackage{balance, url}
\usepackage{multirow, multicol}
\usepackage{booktabs}

\ninept

\title{The NIST CTS Speaker Recognition Challenge}

\makeatletter
\def\name#1{\gdef\@name{#1\\}}
\makeatother
\name{{\em Seyed Omid Sadjadi$^{\textrm{1}}$, Craig Greenberg$^\textrm{1}$, Elliot Singer$^{\textrm{2},\dagger}$,}\\
	{\em Lisa Mason$^\textrm{3}$, Douglas Reynolds$^{\textrm{3}}$} \thanks{$^\dagger$The work of MIT Lincoln Laboratory is supported by the Department of Defense under Air Force Contract FA8702-15-D-0001. Any opinions, findings, conclusions or recommendations expressed in this document are those of the authors do not necessarily reflect the views of the Department of Defense.} }
\address{
	$^\textrm{1}$NIST ITL/IAD/Multimodal Information Group, MD, USA\\
	$^\textrm{2}$MIT Lincoln Laboratory, MA, USA\\
	$^\textrm{3}$U.S. Department of Defense, MD, USA \\
	{\small \tt craig.greenberg@nist.gov} }

\graphicspath{{./figs/}}

\begin{document}
	\maketitle

	\begin{abstract}
		Following the success of the 2019 conversational telephone speech (CTS) speaker recognition challenge, which received 1347 submissions from 67 academic and industrial organizations, the US National Institute of Standards and Technology (NIST) has been conducting a second iteration of the CTS challenge since August 2020. The current iteration of the CTS Challenge is a leaderboard-style speaker recognition evaluation using telephony data extracted from the unexposed portions of the Call My Net 2 (CMN2) and Multi-Language Speech (MLS) corpora collected by the LDC. The CTS Challenge is currently organized in a similar manner to the SRE19 CTS Challenge, offering only an \textit{open} training condition using two evaluation subsets, namely \textit{Progress} and \textit{Test}. Unlike in the SRE19 Challenge, no training or development set was initially released, and NIST has publicly released the leaderboards on both subsets for the CTS Challenge. Which subset (i.e., \textit{Progress} or \textit{Test}) a trial belongs to is unknown to challenge participants, and each system submission needs to contain outputs for all of the trials. The CTS Challenge has also served, and will continue to do so, as a prerequisite for entrance to the regular SREs (such as SRE21). Since August 2020, a total of 53 organizations (forming 33 teams) from academia and industry have participated in the CTS Challenge and submitted more than 4400 valid system outputs. This paper presents an overview of the evaluation and several analyses of system performance for some primary conditions in the CTS Challenge. The CTS Challenge results thus far indicate remarkable improvements in performance due to 1) speaker embeddings extracted using large-scale and complex neural network architectures such as ResNets along with angular margin losses for speaker embedding extraction, 2) extensive data augmentation, 3) the use of large amounts of in-house  proprietary data from a large number of labeled speakers, 4) long-duration fine-tuning. 
	\end{abstract}

	\section{Introduction}
	
	The United States National Institute of Standards and Technology (NIST) has been conducting the second iteration of the conversation telephone speech (CTS) speaker recognition challenge since August 2020. The first iteration of the challenge was organized in 2019~\cite{nistsre19cts}. The CTS Challenge is part of an ongoing series of speaker recognition evaluations (SRE) hosted by NIST since 1996~\cite{nistsre,twodecades}. The objectives of the evaluation series are 1) for NIST to effectively measure system-calibrated performance of the current state of technology, 2) to provide a common test bed that enables the research community to explore promising new ideas in speaker recognition, and 3) to support the community in its development of advanced technology incorporating these ideas. The basic task in the evaluations is speaker detection, that is, determining whether a specified target speaker is talking in a given test speech recording.
	
	The current iteration of the CTS Challenge is a leaderboard-style speaker recognition evaluation using telephony data extracted from the unexposed portions of the Call My Net 2 (CMN2)~\cite{cmn2} (the callee sides only) and Multi-Language Speech (MLS\footnote{This dataset is also known as MLS'14.})~\cite{mls} (the claque/caller sides only) corpora collected by the Linguistic Data Consortium (LDC). The CMN2 (both the caller and callee sides) was previously used to compile the development and test sets for the 2018 and 2019 SREs~\cite{nistsre18_is19, nistsre19cts}, while the MLS (the callee sides only) was used to create the development and test sets for the 2015 and 2017 language recognition evaluations (LRE)~\cite{nistlre15, nistlre17_odyssey18}. This paper describes the task, the performance metric, data, and the evaluation protocol as well as results and performance analyses of submissions received as of December 2021 for the CTS Challenge. The CTS challenge has also served, and will continue to serve, as a prerequisite for the regular SREs (such as SRE19~\cite{nistsre19av} and SRE21~\cite{nistsre21av}), meaning that in order to participate in the regular evaluations, one must first register and make a valid submission for the challenge. The CTS Challenge is coordinated entirely online using an evaluation management platform\footnote{https://sre.nist.gov/cts-challenge} that supports a variety of evaluation-related services such as registration, data license agreement management, data distribution, system output submission and validation/scoring, and system description/presentation uploads.
	
	 The CTS Challenge is organized in a similar manner to the SRE19 CTS Challenge \cite{nistsre19cts}, offering only the \textit{open} training condition in which participants are allowed to use any publicly available and/or proprietary data for system training and development purposes. In addition, similar to the 2019 evaluation, the evaluation set for the CTS Challenge consists of two subsets: a \textit{Progress} subset, and a \textit{Test} subset. Trials for the \textit{Progress} subset comprise approximately 30\% of the target speakers from the unexposed portions of the CMN2 and MLS corpora and is used to monitor progress on the leaderboard, while trials from the remaining 70\% of the speakers are allocated for the \textit{Test} subset. Which subset (i.e., \textit{Progress} or \textit{Test}) a trial belongs to is unknown to challenge participants, and each system submission has to contain outputs for all of the trials.
	
	There are several differences between the SRE19 CTS Challenge and the current iteration; first, unlike in the 2019 evaluation that provided a large in-domain development set, no training or development set was initially released for the CTS Challenge. Nevertheless, in August 2021, i.e., one year after the launch of the CTS Challenge, NIST released the CTS Superset \cite{sadjadi2021nist} which is a large-scale telephony dataset with more than 600,000 segments from nearly 6800 speakers with uniform metadata and keys. Second, information about the data sources used to created the CTS Challenge \textit{Progress} and \textit{Test} sets (i.e., CMN2 and MLS) remained undisclosed to participants until December 2021 to prevent fine-tuning, as well as to increase the difficulty of the evaluation. Finally, NIST publicly displays the leaderboards for both \textit{Progress} and \textit{Test} sets. The \textit{Progress} leaderboard is live and updated after every new submission, while the \textit{Test} leaderboard is updated periodically.  
	
\begin{figure}[!t]
		\centering
		\includegraphics[width=\linewidth, clip, trim=3mm 0mm 0mm 0mm]{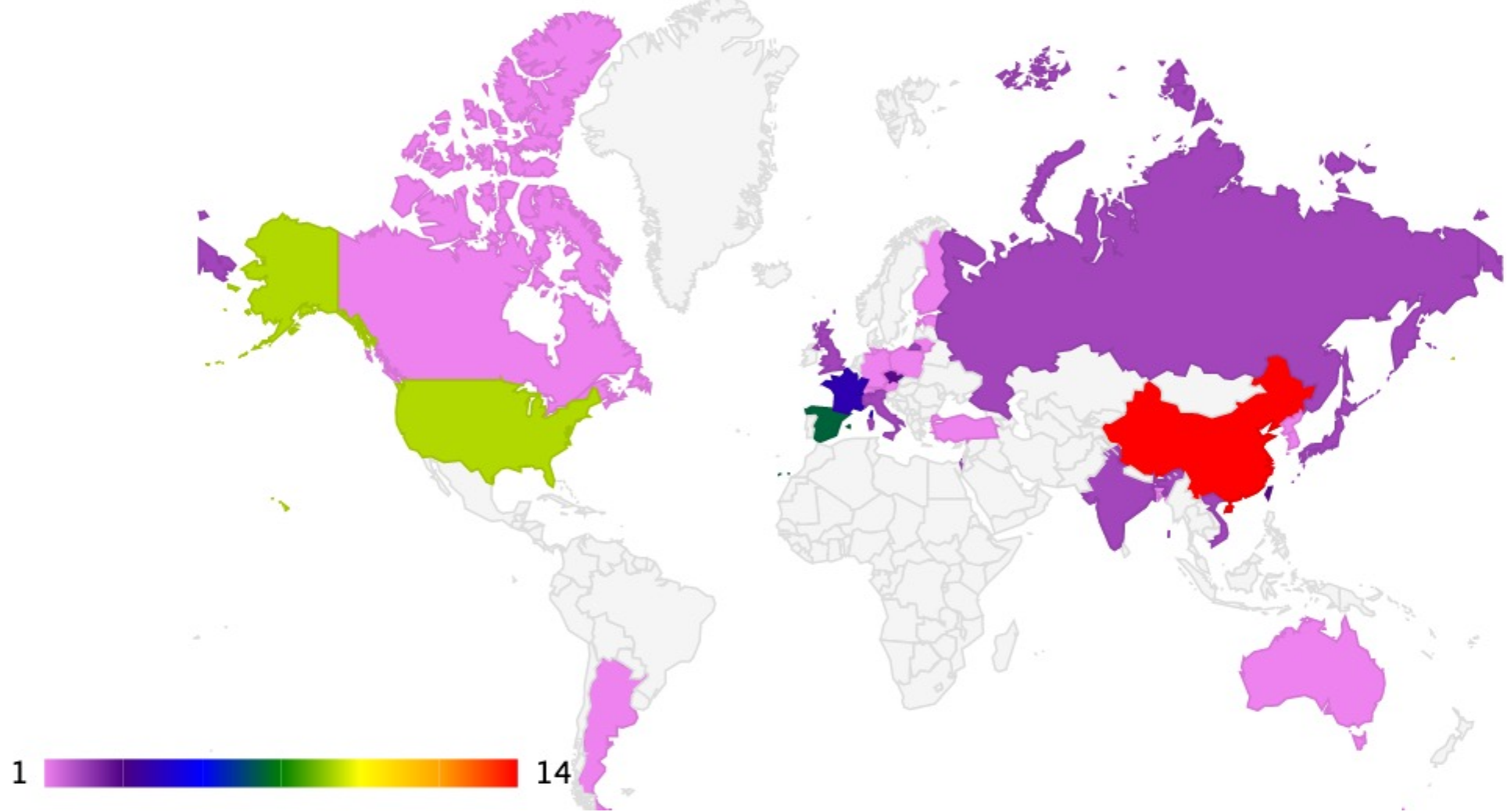}
		\caption{{\it Geographic heatmap for the CTS Challenge participating sites. Colors denote the number of sites per country.}}
		\label{fig:map}
	\end{figure}

\begin{figure}[t]
		\centering
		\includegraphics[width=\linewidth, clip, trim=0mm 10mm 0mm 0mm]{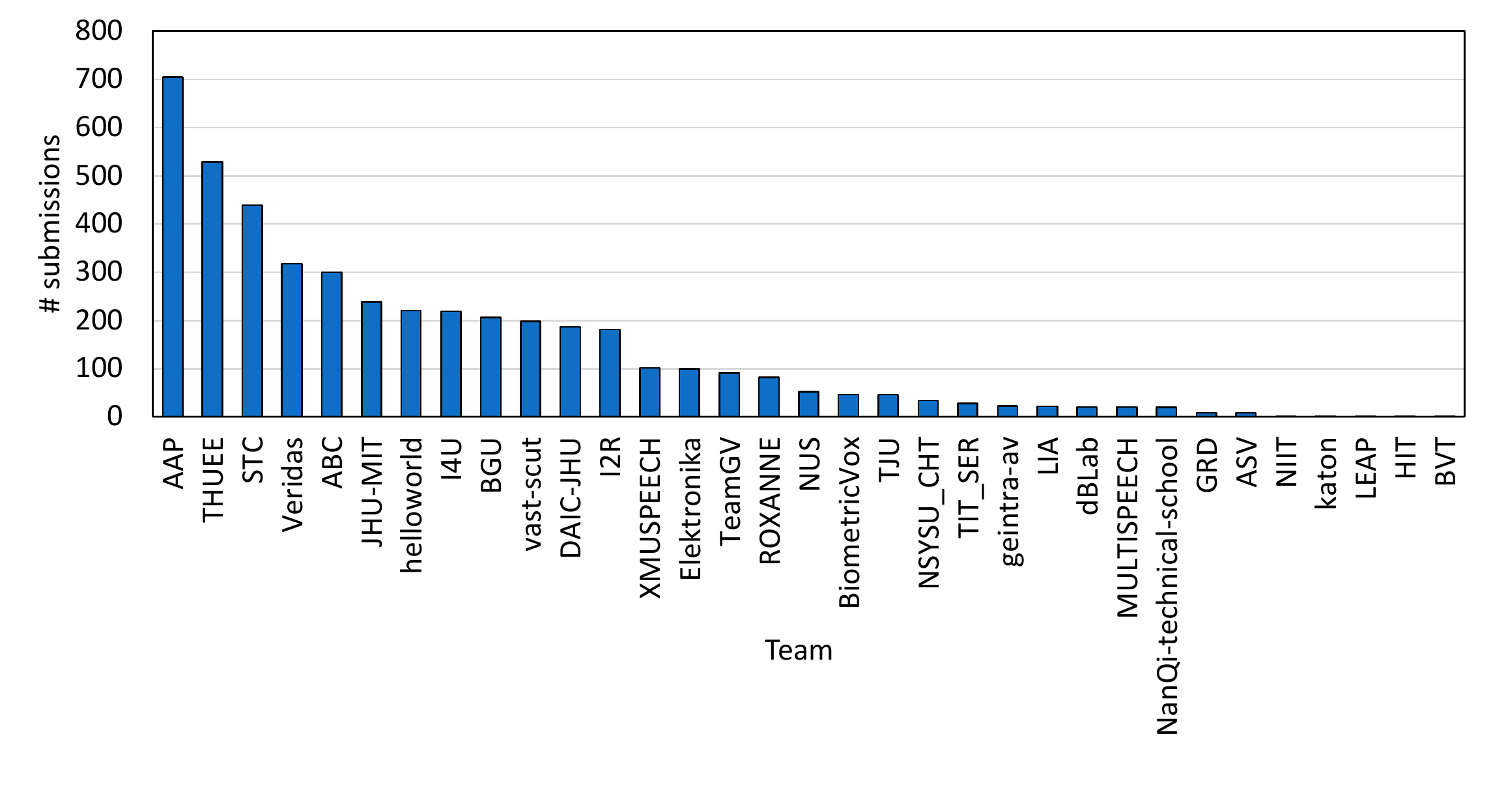}
		\caption{{\it Submission statistics for the CTS Challenge.}}
		\label{fig:submission_stats}
	\end{figure}

	The CTS Challenge participants can make up to 3 submissions per day, and the leaderboard displays the best submission performance results thus far received and processed. Over the course of the challenge since August 2020 through December 2021, a total of 33 teams formed by 53 sites, 23 of which are led by industrial institutions, made more than 4400 valid submissions (note that the participants processed the data locally and submitted only the output of their systems to NIST for scoring and analysis purposes). Figure~\ref{fig:map} displays a geographical heatmap representing the number of participating sites per country. It should be noted that all participant information, including country, are self-reported. The number of submissions per team, as of December 2021, in the CTS Challenge is shown in Figure~\ref{fig:submission_stats}.
	
	Finally, as in the recent SREs, and in an effort to provide a reproducible state-of-the-art baseline for the CTS Challenge, NIST released a report \cite{sadjadi2021nist} containing the baseline speaker recognition system description and results obtained using a state-of-the-art (as of SRE19) deep neural network (DNN) embedding based system (see Section~\ref{sec:baseline} for more details). 
	
	\section{Task Description}
	
	The task for the CTS Challenge is \textit{speaker detection}, meaning given a segment of speech and the target speaker enrollment data, automatically determine whether the target speaker is speaking in the segment. A segment of speech (test segment) along with the enrollment speech segment(s) from a designated target speaker constitute a \textit{trial}. The system is required to process each trial independently and to output a log-likelihood ratio (LLR), using natural (base $e$) logarithm, for that trial. The LLR for a given trial including a test segment $s$ is defined as follows
	
	\begin{equation}
	LLR(s) = \log \left(\frac{P\left(s|H_0\right)}{P\left(s|H_1\right)} \right),
	\end{equation}
	where $P\left(\cdot\right)$ denotes the probability density function (pdf), and $H_0$ and $H_1$ represent the null (i.e., $s$ is spoken by the enrollment speaker) and alternative (i.e., $s$ is not spoken by the enrollment speaker) hypotheses, respectively.
	
	\section{Data}
	\label{sec:data}
	
	In this section we provide a brief description of the data released, as of December 2021, in the CTS Challenge for system training, development, and test.
	
	\subsection{Training set}
	As noted previously, unlike in the SRE19 CTS Challenge that provided large amounts of in-domain speaker-labeled and unlabeled data, no training or development data was initially released for the CTS Challenge. It only offers the \textit{open} training condition that allows the use of any publicly available and/or proprietary data for system training and development purposes. Nevertheless, as of August 2021, NIST has released the NIST SRE CTS Superset (LDC2021E08~\cite{supersetldc}) which is a large-scale dataset for telephony speaker recognition, containing more than 600,000 segments from 6867 speakers with speech durations uniformly distributed in the [10s, 60s] range. The CTS Superset provides uniform metadata and keys for the segments including information on subject ids, session ids, phone ids, gender, language, and duration. For more details about the CTS Superset, we refer the reader to~\cite{sadjadi2021nist}. Although not specified explicitly in the evaluation plan~\cite{sadjadi2020nist}, based on the presentations provided by the participants in December 2021, teams also commonly use prior SRE data (1996-2019) as well as VoxCeleb\footnote{\label{voxceleb}\url{http://www.robots.ox.ac.uk/~vgg/data/voxceleb/}} for system training and development purposes in the CTS Challenge.
	
	
	\subsection{Evaluation set}
	
	\setlength{\tabcolsep}{3mm}
	\renewcommand{\arraystretch}{1.1}
	\begin{table*}[t]
		\caption{\it Data statistics for the CTS Challenge \textit{Progress} and \textit{Test} sets per data source, i.e., CMN2 and MLS.}
		\label{tab:data_stats}
		\centering
		\begin{tabular}{llccccc}
			\toprule
			\multirow{2}{*}{\textbf{Source}} & \textbf{Set} & \textbf{\#speakers} & \textbf{\#1-segment} & \textbf{\#3-segment} & \textbf{\#Test} & \textbf{\#target/non-target} \\
			& & \textbf{(M / F)} & \textbf{enrollment} & \textbf{enrollment} & \textbf{segments} & \textbf{trials} \\
			\midrule
			\multirow{2}{*}{CMN2} & Progress & 25 / 58 & 141 & 29 & 2654 & 1804 / 255,178 \\
			& Test & 61 / 137 & 308 & 55 & 2654 & 4123 / 580,256 \\
			\hline
			\multirow{2}{*}{MLS} & Progress & 22 / 25 & 141 & 47 & 12,249 & 17,992 / 141,584 \\
			& Test& 48 / 81 & 387 & 129 & 17,769 & 53,084 / 351,912 \\
			\bottomrule
		\end{tabular}
	\end{table*}
	
	The speech segments in the CTS Challenge evaluation subsets were extracted from the unexposed portions of the CMN2~\cite{cmn2} (the callee sides only) and MLS~\cite{mls} (the claque/caller sides only) corpora collected by the LDC. The CMN2 corpus, which was previously used for the 2018 and 2019 SREs, consists of CTS recordings spoken in Tunisian Arabic which were collected over the traditional Public Switched Telephone Network (PSTN) and the more recent Voice over IP (VOIP) platforms outside North America. For CMN2 data collection, the LDC recruited a few hundred speakers called \textit{claques} who made multiple calls to people in their social network (e.g., family and friends). Claques were  encouraged to use different telephone instruments (e.g., cell phone, landline) in a variety of settings (e.g., noisy cafe, quiet office) for their initiated calls and were instructed to talk for at least 8--10 minutes on a topic of their choice. All CMN2 recordings are encoded as a-law sampled at 8~kHz in NIST SPHERE \cite{sphere2012} formatted files. For the CTS Challenge, the unexposed callee (i.e., non-claque) sides were used to created the enrollment and test segments. Because the callee sides were neither audited nor assigned speaker labels, we used the phone numbers associated with the calls as pseudo-speaker labels. These pseudo labels were then refined using a speaker recognition system augmented with extensive manual listening to 1) prune calls that did not originate from the pseudo-speakers, 2) split a pseudo-speaker label into multiple speaker labels based on the affinity/similarity matrix computed for the calls associated with that pseudo-speaker, and 3) merge the calls from multiple phone numbers that originated from the same speaker. In order to increase the accuracy of the label refinement process, we used a speaker recognition system trained on in-domain speaker-labeled data from the claque sides of the phone conversations.

	\begin{table*}[b!]
		\caption{\it MLS languages and language clusters.}
		\label{tab:mls_languages}
		\centering{
		\begin{tabular}{|l|l|l|}
			\hline
			\textbf{Cluster (6)} & \textbf{Language (20)} & \textbf{Language code}\\
			\hline \hline
			Arabic & Egyptian, Iraqi, Levantine, Maghrebi, MSA$^+$ & ara-arz, ara-acm, ara-apc, ara-ary, ara-arb \\
			\hline
			Chinese &  Mandarin, Min Nan,  Wu, Cantonese & zho-cmn, zho-nan, zho-wuu, zho-yue \\
			\hline
			English & British$^+$, General American, Indian & eng-gbr, eng-usg, eng-sas \\
			\hline
			Slavic & Polish, Russian & qsl-pol, qsl-rus \\
			\hline
			\multirow{ 2}{*}{Iberian} & Brazilian Portuguese, Caribbean Spanish & por-brz, spa-car, spa-eur, spa-lac \\
			& European Spanish, Latin American Continental Spanish &  \\
			\hline
			French & Haitian, West African & fre-hat, fre-waf \\
			\hline
		\end{tabular}}
		\\ \footnotesize{$^+$No CTS data are available for these languages.}\\
	\end{table*}
	
	The MLS corpus, which was previously used for the 2015 and 2017 LREs, consists of CTS and broadcast narrow-band speech (BNBS) data collected in the U.S. by the LDC. The speech data is spoken in 20 languages representing 6 clusters of confusable varieties (see Table~\ref{tab:mls_languages} for more details). For MLS data collection, the LDC recruited a few hundred \textit{claques} who made multiple calls to people in their social network both inside and outside the U.S. The minimum number of calls per claque is 1, and maximum number of calls is 133. All MLS recordings are encoded as $\mu$-law (8-bit) sampled at 8~kHz in NIST SPHERE \cite{sphere2012} formatted files. We used the unexposed claque sides from the MLS to created enrollment and test segments for the CTS Challenge. Although the claques were assigned unique identifiers (PIN) at the time of collection, they were not audited by the LDC to ensure that each PIN was in fact used by a single speaker. Therefore, we used the PINs as initial speaker labels and refined them using a speaker recognition system augmented with extensive manual listening to achieve the three objectives noted above for the CMN2. In order to increase the accuracy of the label refinement process, we used a speaker recognition system adapted to in-domain data from the callee sides of the MLS.
	
	As noted previously, the CTS Challenge evaluation set consists of a \textit{Progress} set and a \textit{Test} set. Trials for the \textit{Progress} subset comprise approximately 30\% of the target speakers from the the CMN2 and MLS corpora and are used to monitor progress on the live leaderboard, while trials from the remaining 70\% of the speakers are allocated for the \textit{Test} subset. From a total 18 languages in the MLS corpus with CTS data, 8 languages, each with more than 10 speakers, are used both in the \textit{Progress} and \textit{Test} sets, while the remaining 10 (with less than 10 speaker per language) are only used in the \textit{Test} set. The evaluation set (LDC2020E28~\cite{ldc2020e28}) is available through the challenge web platform (\url{https://sre.nist.gov}) subject to the approval of the LDC data license agreement. Table~\ref{tab:data_stats} summarizes the data statistics for CTS Challenge \textit{Progress} and \textit{Test} sets per data source (i.e., CMN2 and MLS). It is worth noting here that the gender labels for the speakers and their corresponding segments were automatically derived using a time-delay neural network (TDNN) based gender identification system. The gender labels for speakers with inconsistent segment-level labels were manually corrected through listening. 
	
	Similar to the most recent SREs, there were two enrollment scenarios in the CTS Challenge, namely 1-segment and 3-segment conditions. As the names imply, in the 1-segment condition only one approximately 60~s speech segment is available for enrollment, while in the 3-segment condition three approximately 60~s speech segments (from the same phone number) are provided to build the model of the target speaker.
	
	\setlength{\tabcolsep}{1.2mm}
	\renewcommand{\arraystretch}{1.}
	\begin{table}[t]
		\caption{\it Primary partitions in the CTS Challenge CMN2 subset}
		\label{tab:primary_partitions_cmn2}
		\centering
		\begin{tabular}{l|llcc}
			\toprule
			\textbf{Subset} & \textbf{Partition} & \textbf{Elements} & \textbf{\#target} & \textbf{\#non-target} \\
			\midrule
			\multirow{4}{*}{Progress} & \multirow{2}{*}{Gender} & male & 501 & 40,552 \\
			& & female & 1303 & 214,626 \\
			\cline{2-5}
			& \multirow{2}{*}{\#enrollment} & 1 & 1461 & 213,768 \\
			& segments  & 3 & 343 & 41,410 \\
			\hline
			\multirow{4}{*}{Test} & \multirow{2}{*}{Gender} & male & 898 & 68,421 \\
			& & female & 3225 & 511,835 \\
			\cline{2-5}
			& \multirow{2}{*}{\#enrollment} & 1 & 3375 & 490,361 \\
			& segments  & 3 & 748 & 89,895 \\
			\bottomrule
		\end{tabular}
	\end{table}
	
	The CTS Challenge test conditions are as follows:
	
	\begin{itemize}
		\item The speech durations of the test segments are uniformly sampled ranging approximately from 10 seconds to 60 seconds. 
		
		\item Trials are conducted with test segments from both same and different phone numbers as the enrollment segment(s).
		
		\item There are no cross-lingual trials.
		
		\item There are no cross-gender trials.
	\end{itemize}
	
	\section{Performance Measurement}
	\label{sec:metric}
	
	\setlength{\tabcolsep}{1.2mm}
	\renewcommand{\arraystretch}{1.}
	\begin{table}[t]
		\caption{\it Primary partitions in the CTS Challenge MLS subset}
		\label{tab:primary_partitions_mls}
		\centering
		\begin{tabular}{l|llcc}
			\toprule
			\textbf{Subset} & \textbf{Partition} & \textbf{Elements} & \textbf{\#target} & \textbf{\#non-target} \\
			\midrule
			\multirow{4}{*}{Progress} & \multirow{2}{*}{Gender} & male & 8668 & 52,300 \\
			& & female & 9324 & 89,284 \\
			\cline{2-5}
			& \multirow{2}{*}{\#enrollment} & 1 & 13,494 & 106,188 \\
			& segments  & 3 & 4498 & 35,396 \\
			\hline
			\multirow{4}{*}{Test} & \multirow{2}{*}{Gender} & male & 18,904 & 119,008 \\
			& & female & 34,180 & 232,904 \\
			\cline{2-5}
			& \multirow{2}{*}{\#enrollment} & 1 & 39,813 & 263,934 \\
			& segments  & 3 & 13,271 & 87,978 \\
			\bottomrule
		\end{tabular}
		\vspace{-4mm}
	\end{table}

	The primary performance measure for the CTS Challenge is a detection cost defined as a weighted sum of false-reject (miss) and false-accept (false-alarm) error probabilities. Equation (\ref{eq: cdet}) specifies the CTS Challenge primary normalized cost function for some decision threshold $\theta$,
	
	\begin{equation} \label{eq: cdet}
	C_{norm}\left(\theta\right) = P_{miss}\left(\theta\right) + \beta \times P_{fa}\left(\theta\right) ,
	\end{equation}
	where $\beta$ is defined as
	\begin{equation}
	\beta = \frac{C_{fa}}{C_{miss}} \times \frac{1-P_{target}}{P_{target}}.
	\end{equation}
	The parameters $C_{miss}$ and $C_{fa}$ are the cost of a missed detection and cost of a false-alarm, respectively, and $P_{target}$ is the \textit{a priori} probability that the test segment speaker is the specified target speaker. The primary cost metric, $C_{primary}$ for the CTS Challenge is the normalized cost calculated at a single operating point along the detection error trade-off (DET) curve \cite{nist1997}, with $C_{miss}=C_{fa}=1$, $P_{target}=0.05$. Here, $\log(\beta)$ is applied as the detection threshold $\theta$ for computing the actual detection costs. Additional details can be found in the CTS Challenge evaluation plan \cite{nist2019_cts}. Two separate normalized costs are calculated for the CMN2 and MLS trials, and the average of the two costs is used as the final metric.

	Similar to the SRE19 CTS Challenge, the evaluation trials are divided into several, but fewer, partitions. Each partition is defined as a combination of the number of enrollment segments (1 vs 3) and speaker gender (male vs female). More information about the various partitions in the CTS Challenge \textit{Progress} and \textit{Test} subsets can be found in Tables~\ref{tab:primary_partitions_cmn2} and \ref{tab:primary_partitions_mls}. An actual $C_{Primary}$ is calculated for each partition, and the final result is a weighted average of actual $C_{Primary}$'s across the various data sources. 

    Also, two minimum detection costs (minimum $C_{Primary}$) are computed (one per data source) by using the detection thresholds that minimize the detection costs for each data source (i.e., one for CMN2 and another for MLS). The minimum $C_{Primary}$'s are then averaged. Note that for minimum cost calculations, the counts for each condition set are equalized before pooling and cost calculation (i.e., minimum cost is computed using a single threshold, not one per condition set).

	\section{Baseline System}
	\label{sec:baseline}
	\begin{figure}[t]
		\centering
		\includegraphics[width=\linewidth, clip, trim=70mm 80mm 50mm 50mm]{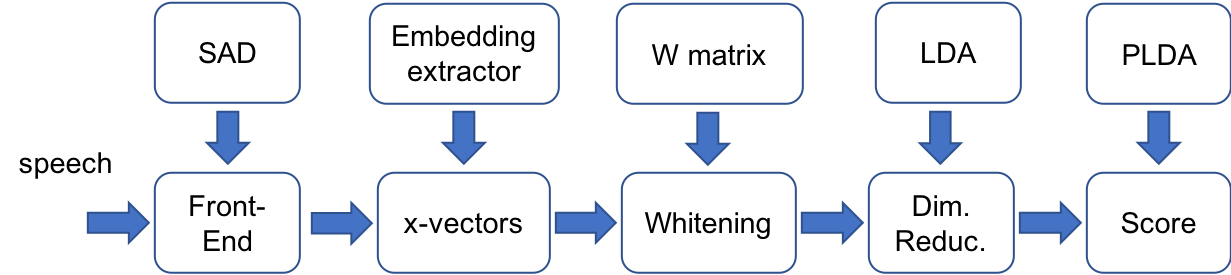}
		\caption{{\it A simplified block diagram of the baseline speaker recognition system for the CTS Challenge.}}
		\label{fig:blockdiag_xvec}
		\vspace{-4mm}
	\end{figure}
	
	In this section, we describe the baseline speaker recognition system setup including speech and non-speech data used for training the system components as well as the hyper-parameter configurations used. Figure~\ref{fig:blockdiag_xvec} shows a simplified block diagram of the x-vector baseline system. The embedding extractor is trained using Pytorch\footnote{https://github.com/pytorch/pytorch}, while the NIST SLRE toolkit is used for front-end processing and back-end scoring.

    \subsection{Data}

    The baseline system is developed using the CTS Superset (LDC2021E08~\cite{supersetldc}) described in~\cite{sadjadi2021nist}. In order to increase the diversity of the acoustic conditions in the training set, two different data augmentation strategies are adopted. The first strategy uses noise-degraded (using babble, general noise, and music) versions of the original recordings, while the second strategy uses spectro-temporal masking applied directly on spectrograms (aka spectrogram augmentation \cite{specaug}). The noise samples for the first augmentation approach are extracted from the MUSAN corpus \cite{musan}. For spectrogram augmentation, the mild and strong policies described in \cite{specaug} are used.

    \subsection{Configuration}

    For speech parameterization, we extract 64-dimensional log-mel spectrograms from 25 ms frames every 10 ms using a 64-channel mel-scale filterbank spanning the frequency range 80~Hz--3800~Hz. After dropping the non-speech frames using SAD, a short-time cepstral mean subtraction is applied over a 3-second sliding window.

    For embedding extraction, an extended TDNN \cite{snyder2019} with 11 hidden layers and parametric rectified linear unit (PReLU) non-linearities is trained to discriminate among the nearly 6800 speakers in the CTS Superset set. A cosine loss with additive margin \cite{cosface} is used in the output layer (with $m=0.2$ and $s=40$). The first 9 hidden layers operate at frame-level, while the last 2 operate at segment-level. There is a 3000-dimensional statistics pooling layer between the frame-level and segment-level layers that accumulates all frame-level outputs from the 9\textsuperscript{th} layer and computes the mean and standard deviation over all frames for an input segment. The model is trained using Pytorch and the stochastic gradient descent (SGD) optimizer with momentum ($0.9$), an initial learning rate of $10^{-1}$, and a batch size of $512$. The learning rate remains constant for the first $5$ epochs, after which it is halved every other epoch. 

    To train the network, a speaker-balanced sampling strategy is implemented where in each batch 512 unique speakers are selected, without replacement, from the pool of training speakers. Then, for each speaker, a random speech segment is selected from which a 400-frame (corresponding to 4 seconds) chunk is extracted for training. This process is repeated until the training samples are exhausted.

    After training, embeddings are extracted from the 512-dimensional affine component of the 10\textsuperscript{th} layer (i.e., the first segment-level layer). Prior to dimensionality reduction through linear discriminant analysis (LDA) to 250, 512-dimensional embeddings are centered, whitened, and unit-length normalized. The centering and whitening statistics are computed using the CTS Superset data. For backend scoring, a Gaussian probabilistic LDA (PLDA) model with a full-rank Eigenvoice subspace is trained using the embeddings extracted from  only the original (as opposed to degraded) speech segments in the CTS Superset. No parameter/domain adaptation or score normalization/calibration is applied.

	\begin{figure}[t]
		\centering
		\includegraphics[width=\linewidth, clip, trim=0mm 22mm 2mm 0mm]{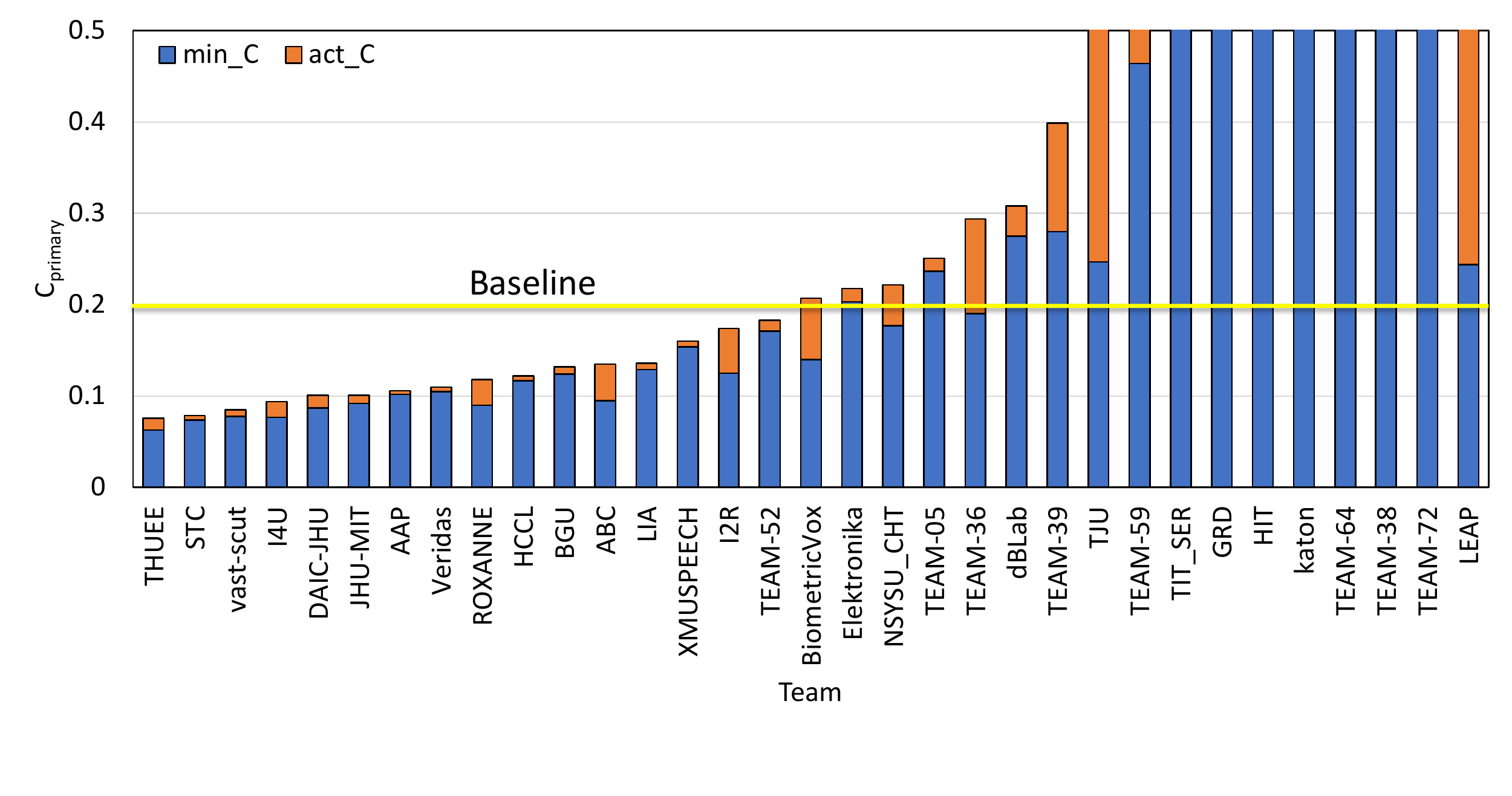}
		\vspace{-4mm}
		\includegraphics[width=\linewidth, clip, trim=0mm 0mm 0mm 0mm]{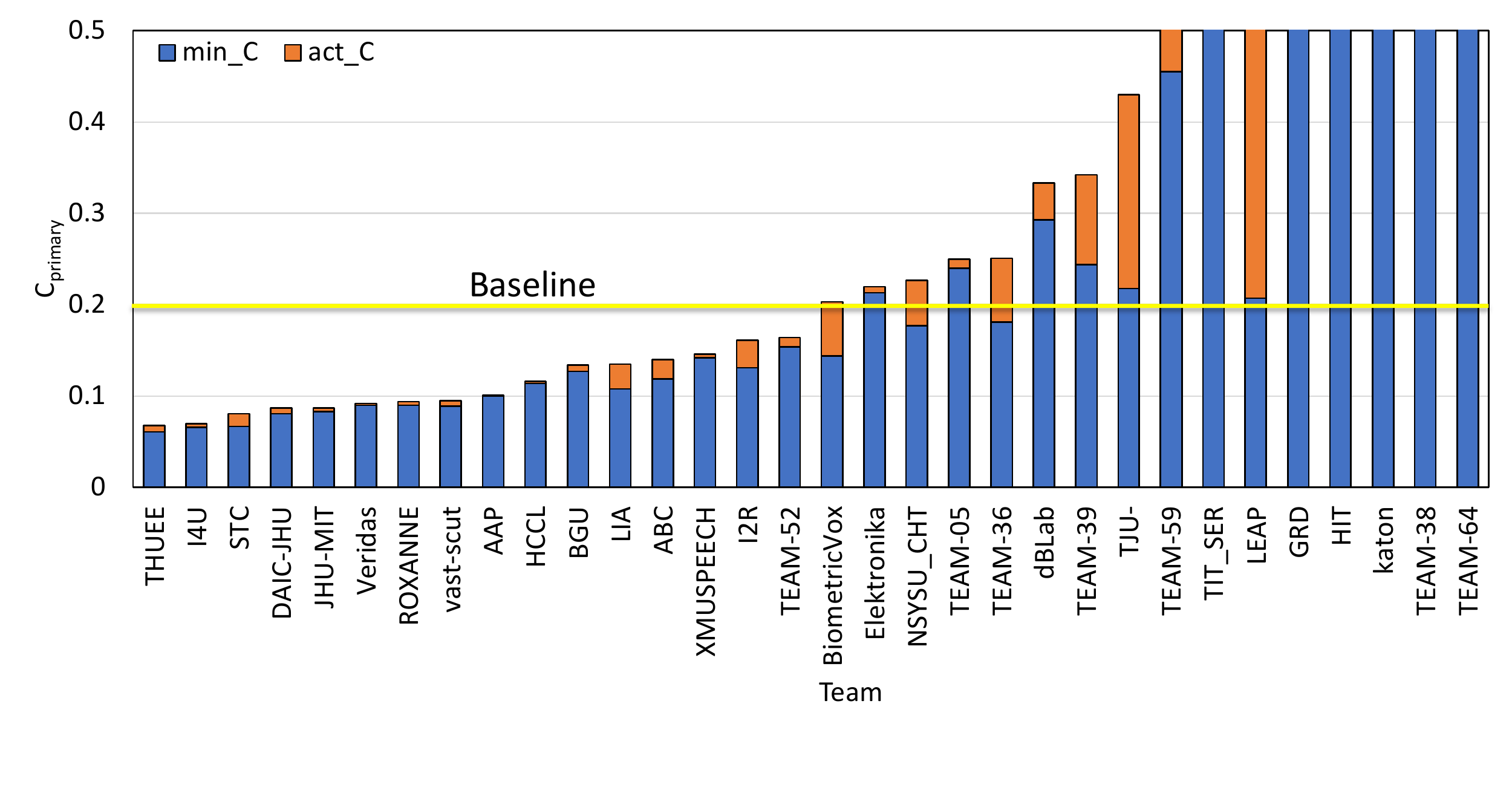}
		\vspace{-6mm}
		\caption{\it Performance of the CTS Challenge submissions in terms of actual (orange) and minimum (blue) costs for the \textit{Progress} (top) and \textit{Test} (bottom) subsets.}
		\label{fig:cts_progress_test}
		\vspace{-4mm}
	\end{figure}

	\begin{figure}[t]
		\centering
		\includegraphics[width=\linewidth, height=3.5cm, clip, trim=0mm 10mm 0mm 0mm]{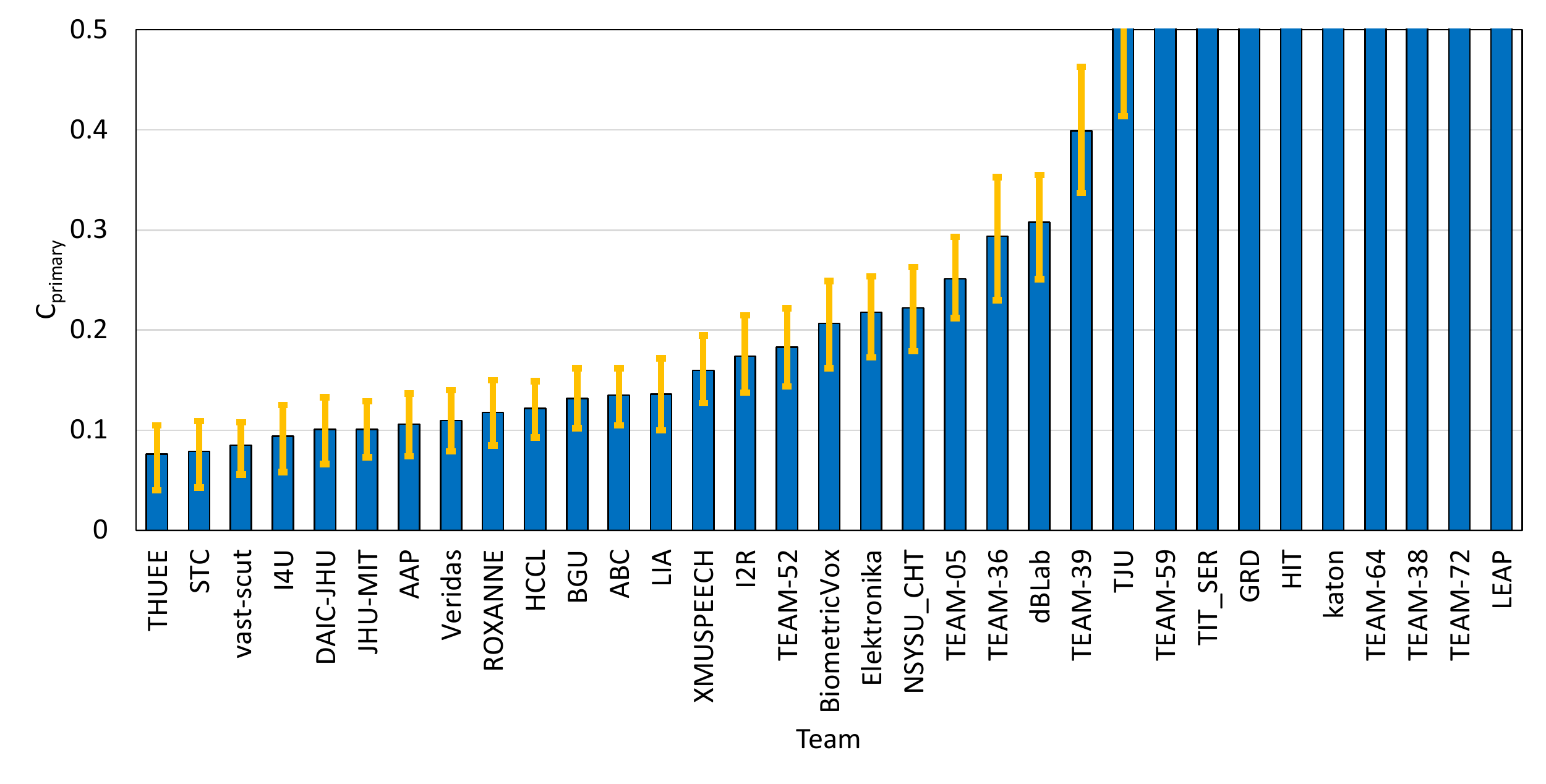}
		\vspace{-4mm}
		\includegraphics[width=\linewidth, height=3.8cm, clip, trim=0mm 0mm 0mm 0mm]{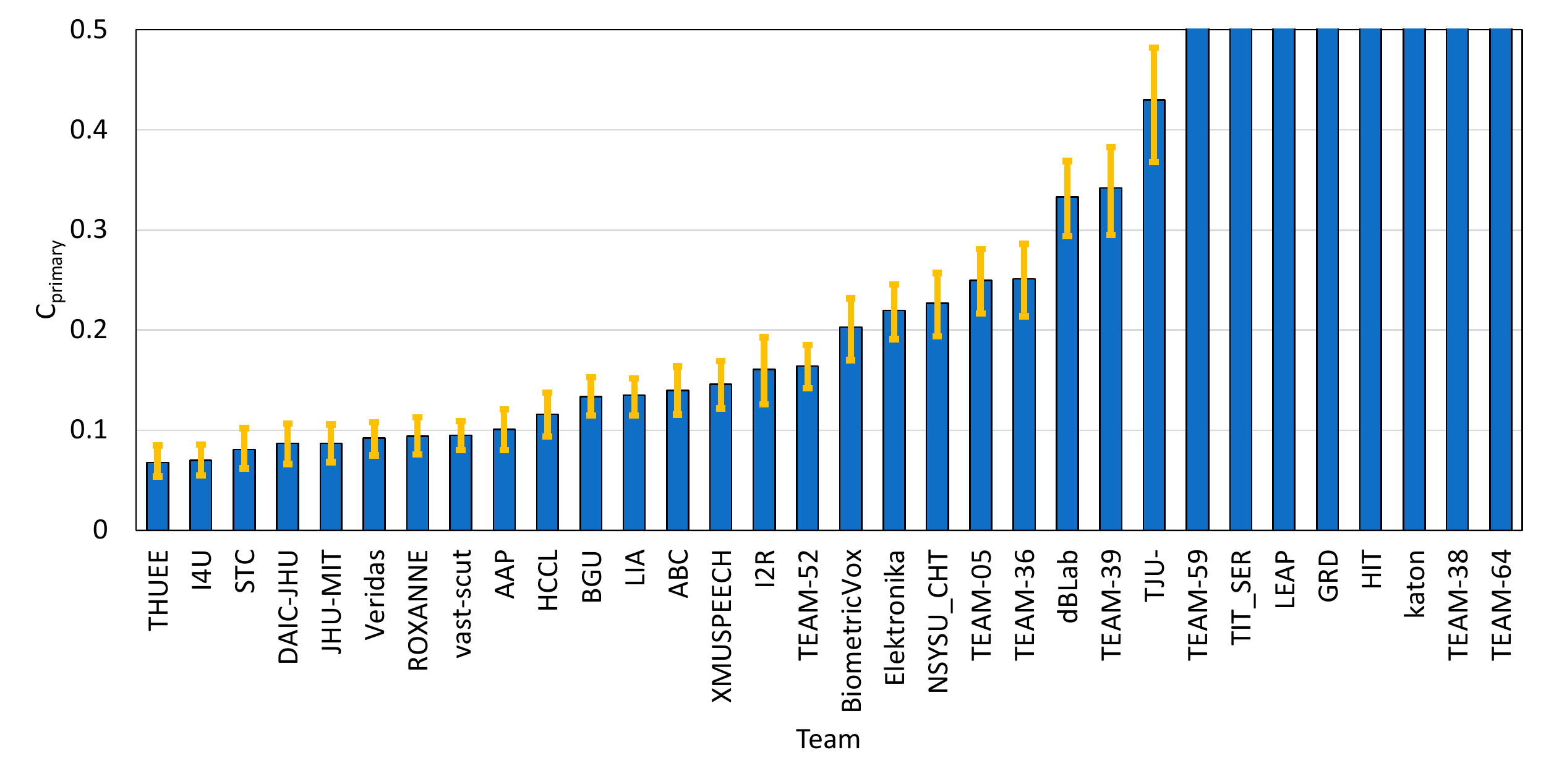}
		\vspace{-2mm}
		\caption{\it Performance confidence intervals (95\%) of the CTS Challenge submissions for the \textit{Progress} (top) and \textit{Test} (bottom) subsets.}
		\label{fig:cts_progress_test_ci}
		\vspace{-4mm}
	\end{figure}
	
	\section{System Highlights}
	
	Before presenting the results and performance analyses, it is worthwhile to briefly review some highlights of the top performing systems in the CTS Challenge as of December 2021. In terms data usage, a majority of the systems used data extracted from prior SREs along with the Voxceleb corpus. A few systems also utilized the data from prior LREs~\cite{nistlre17_odyssey18}, while a few others used in-house datasets, some including recordings from more than 200k speakers. The top performing systems used extensive data augmentation to increase the diversity of the available data. In terms of core system components, all teams used neural embeddings as speaker representations; the embeddings were extracted from different variations of ResNets~\cite{he2016deep} trained using different flavors of angular margin losses (e.g., \cite{cosface, arcface}). Top performers further used domain-dependent training, long duration fine-tuning~\cite{romero2019fine}, along with adaptive score normalization~\cite{cumani2011}. For system combination, teams used early or/and late fusion strategies. Finally, we saw a single best system focus for majority of the systems.
	
	\section{Results and Discussion}
	
	In this section we present some key results and analyses for the CTS Challenge, in terms of minimum and actual costs as well as DET performance curves.
	
	Figure~\ref{fig:cts_progress_test} shows performance of the best submissions per team per subset in terms of the actual (orange) and minimum (blue) costs, for the CTS Challenge \textit{Progress} and \textit{Test} subsets, respectively. The yellow horizontal line denotes the performance of the baseline systems in terms of minimum cost (see Section~\ref{sec:baseline} for more details). Here, the y-axis limit is set to $0.5$ to facilitate cross-system comparisons in the lower cost region. Several observations can be made from the two plots. First, performance trends on the two subsets are generally similar, although slightly better results are observed on the \textit{Test} subset compared to the \textit{Progress} subset. We speculate that this rather counter-intuitive phenomenon results primarily from the inclusion of languages with small number speakers (i.e., less than 10 speakers) in the \textit{Test} set. Note that there no cross-lingual trials in the CTS Challenge. Second, nearly half of the submissions outperform the baseline system trained on the CTS Superset, with the top performers achieving more than 50\% improvements over the baseline. Third, a majority of the systems achieve relatively small calibration errors on both the \textit{Progress} and \textit{Test} subsets, despite the fact that the data sources for the CTS Challenge were undisclosed. The good calibration performance is in line with that of the submitted systems for the SRE19 CTS Challenge. Finally, it can be seen from the figures that the performance gap among the top-5 teams is not remarkable. We will look into the actual performance differences among the top performing teams from a statistical significance perspective next.

    \begin{figure}[t]
		\centering
		\includegraphics[scale=0.8, clip, trim=0mm 0mm 5mm 7mm]{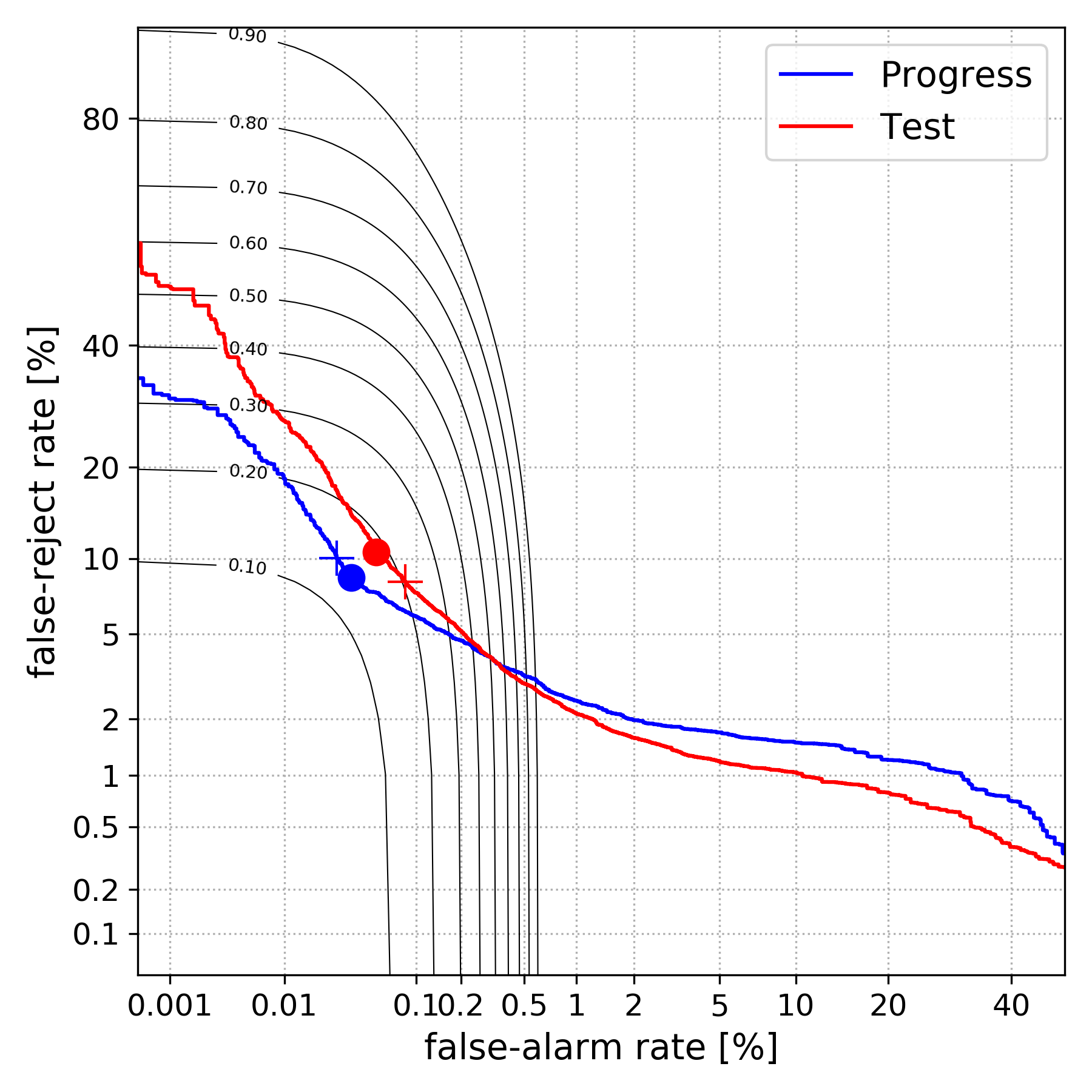}
		\vspace{-4mm}
		\caption{\it DET performance curves of a leading system for the \textit{Progress} and \textit{Test} subsets in CMN2 (top) and MLS (bottom) trials.} 
		\label{fig:progress_vs_test}
		\vspace{-4mm}
	\end{figure}
	
	\begin{figure}[t]
		\centering
		\includegraphics[scale=0.45, clip, trim=0mm 0mm 5mm 10mm]{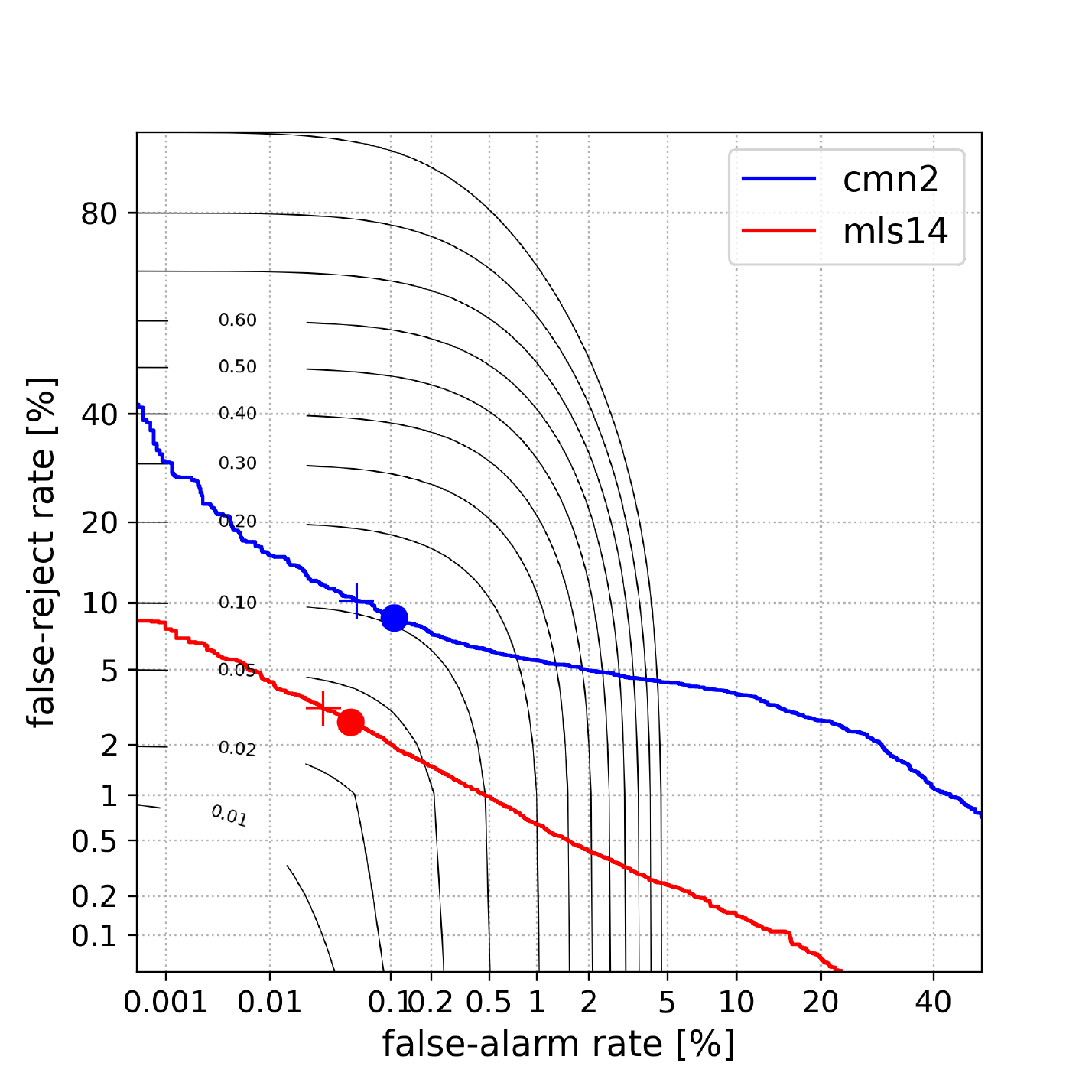}
		\vspace{-4mm}
		\caption{\it DET performance curves of a leading system as a function of data source (CMN2 vs MLS).} 
		\label{fig:cmn2_vs_mls}
		\vspace{-4mm}
	\end{figure}
	
	\begin{figure}[t]
		\centering
		\includegraphics[scale=0.45, clip, trim=0mm 0mm 0mm 0mm]{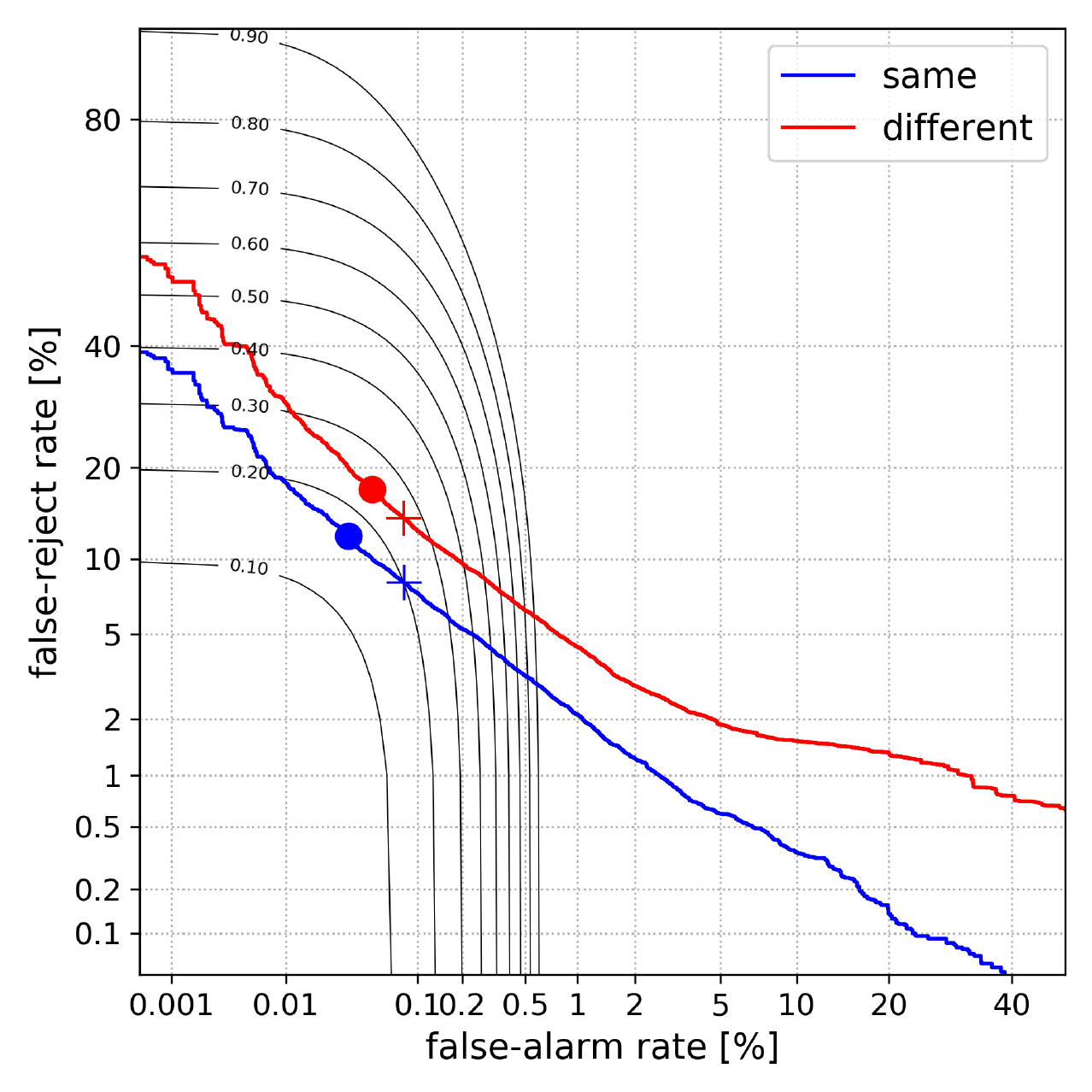}
		\vspace{-4mm}
		\caption{\it DET performance curves of a leading system as a function of enrollment-test phone number match (same vs different) for CMN2 trials.} 
		\label{fig:phone_num_match}
		\vspace{-4mm}
	\end{figure}
	
	\begin{figure}[t]
		\centering
		\includegraphics[scale=0.75, clip, trim=0mm 0mm 0mm 0mm]{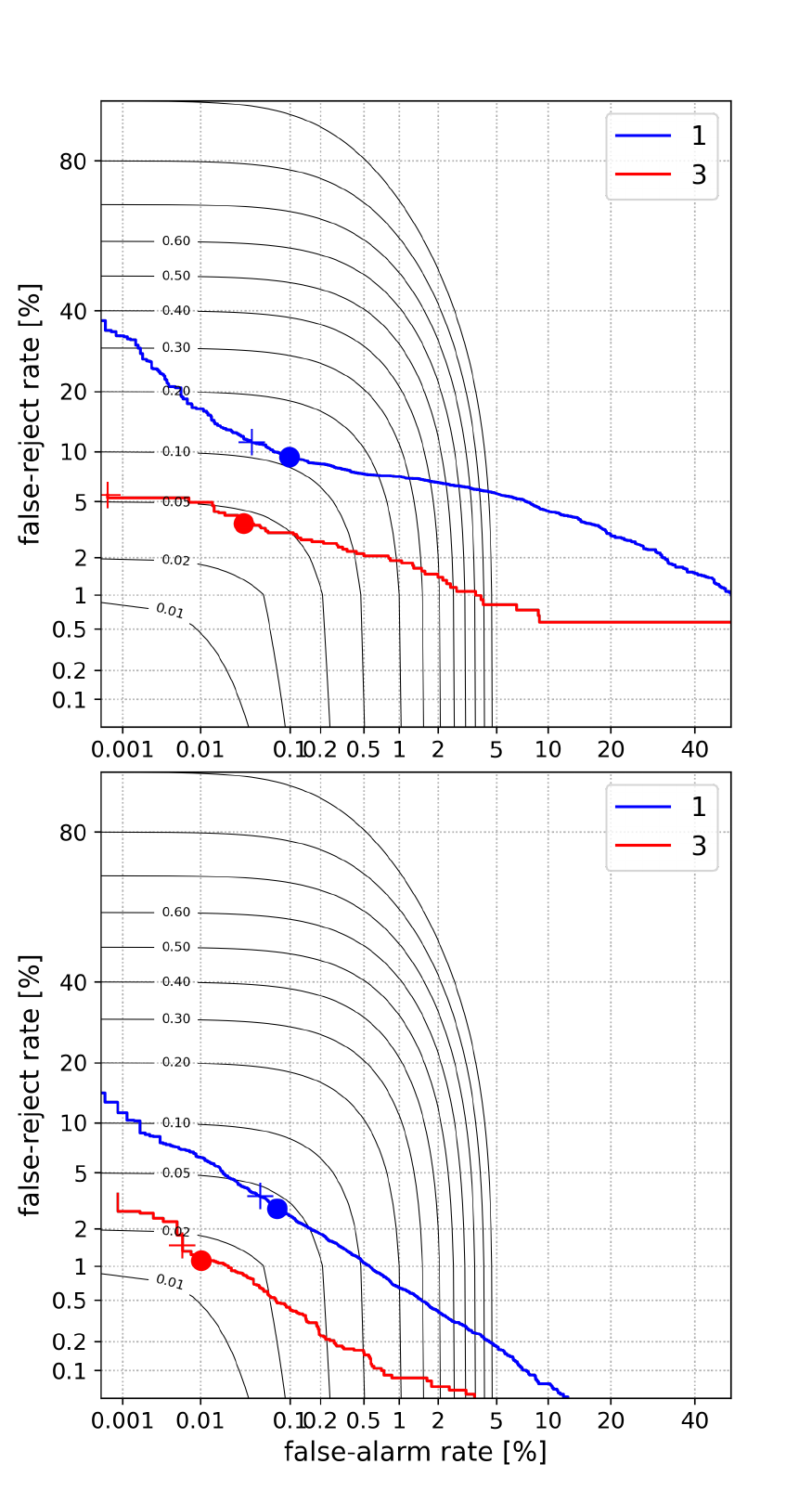}
		\vspace{-4mm}
		\caption{\it DET performance curves of a leading system as a function of the number of enrollment segments for CMN2 (top) and MLS (bottom) trials.} 
		\label{fig:num_enroll_segs}
		\vspace{-3mm}
	\end{figure}

	\begin{figure}[t]
		\centering
		\includegraphics[scale=0.75, clip, trim=0mm 0mm 0mm 0mm]{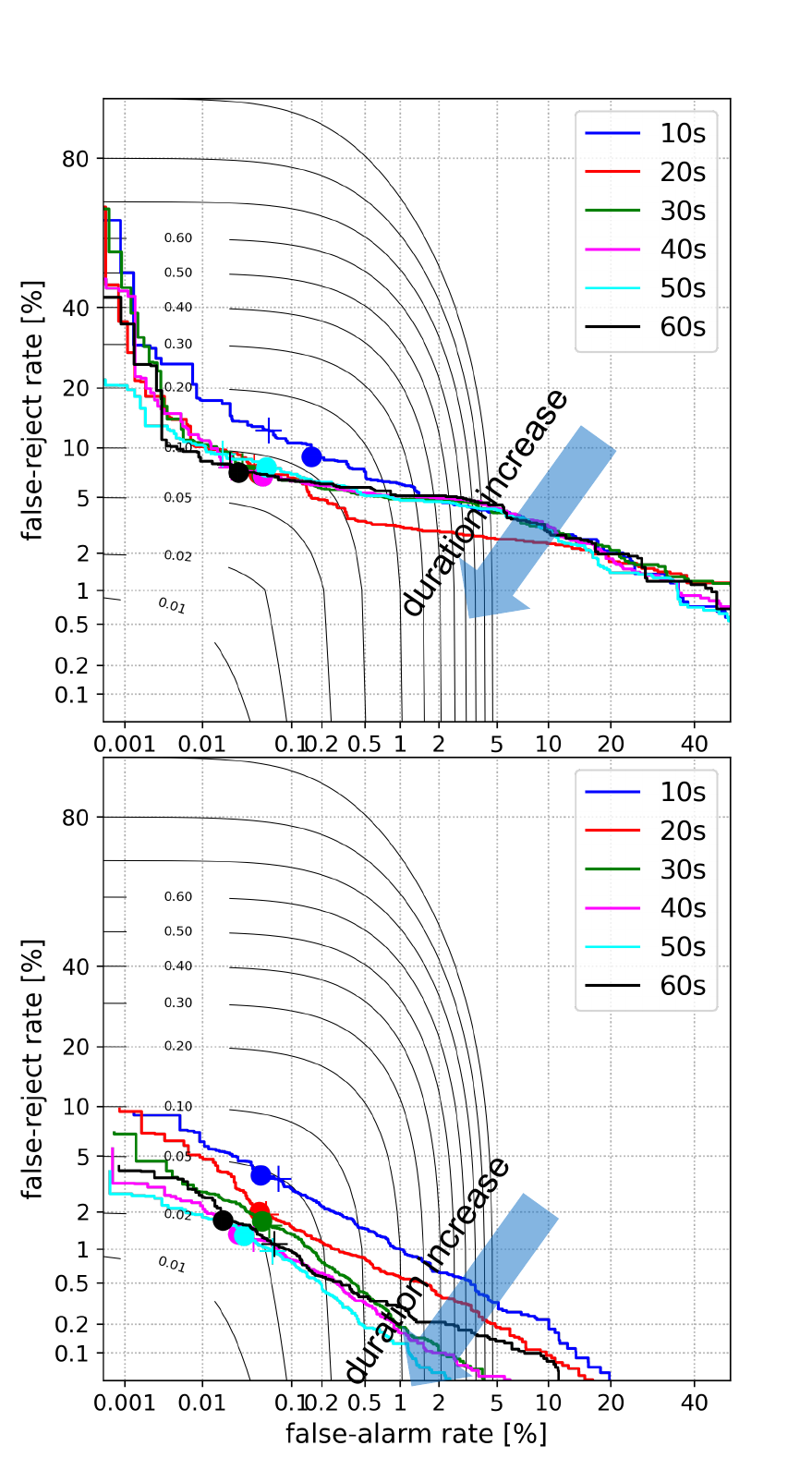}
		\vspace{-4mm}
		\caption{\it DET curve performances of a top performing system for the various segment speech durations (10~s--60~s) in the test set, for CMN2 (top) and MLS (bottom) trials.}
		\label{fig:duration}
		\vspace{-3mm}
	\end{figure}
	
	It is common practice in the machine learning community to perform statistical significance tests to facilitate a more meaningful cross-system performance comparison. Accordingly, similar to our SRE19 performance analysis, and to encourage the speaker recognition community to consider significance testing while comparing systems or performing model selection, we computed bootstrapping-based 95\% confidence intervals using the approach described in \cite{bootstrap}. To achieve this, we sampled, with repetition, the unique speaker model space along with the associated test segments 1000 times, which resulted in 1000 actual detection costs, based on which we calculated the quantiles corresponding to the 95\% confidence margin. Figure~\ref{fig:cts_progress_test_ci} shows the performance confidence intervals (around the actual detection costs) for each submission for both the \textit{Progress} (top) and \textit{Test} (bottom) subsets. It can be seen that, in general, the \textit{Progress} subset exhibits a wider confidence margin than the \textit{Test} subset, which is expected because it has a relatively smaller number of trials. Also, notice that a majority of the top systems perform comparably under different samplings of the trial space. It is also interesting to observe that for some top-performing systems the confidence intervals are much narrower compared to the others which could indicate robustness to the various samplings of the trial space. These observations further highlight the importance of statistical significance tests while reporting performance results or in the model selection stage during system development, in particular when the number of trials is relatively small.
	
	Figures~\ref{fig:progress_vs_test} shows DET curves for a top performing submission as a function of evaluation subset (i.e., \textit{Progress} vs \textit{Test}), separately for CMN2 and MLS trials. The closer the curves to the origin, the better the performance. The circle and crosses denote the minimum and actual costs, while the solid black curves represent the equi-cost contours, meaning that all points are on each curve correspond to the same cost value. Firstly, consistent with our observations from Figure~\ref{fig:cts_progress_test}, the detection errors (i.e., false-alarm and false-reject errors) across the operating points of interest (i.e., the low false-alarm region) for the \textit{Test} subset are slightly smaller than those for the \textit{Progress} subset, in particular for the MLS trials. This could be attributed to the inclusion of languages with small numbers of speakers in the \textit{Test} subset for the MLS, although this only remains a hypothesis. In addition, depending on the data source (i.e., CMN2 or MLS) the calibration error for the \textit{Test} subset is either larger or smaller than that for the \textit{Progress} subset.
	
	Figure~\ref{fig:cmn2_vs_mls} highlights the performance differences of a top performing system on CMN2 versus MLS, where drastically better performance is observed on the MLS. Again, this is primarily attributed to the relatively small number of speakers within each language class for the MLS, where the maximum number of speakers per language is 20. It is worth re-emphasizing that there are no cross-lingual trials in the CTS Challenge. On the contrary, there are nearly 200 speakers in the CMN2 \textit{Test} subset all speaking the same language (i.e., Tunisian Arabic).
	
	The impact of enrollment-test phone number match is shown in Figure~\ref{fig:phone_num_match}, only for CMN2. Note that the phone number information was not available for MLS. Consistent with the results from our previous evaluations, we see noticeably better performance when enrollment and test segments originate from the same phone number. Nevertheless, the error rates still remain relatively high for the same phone number condition, pointing to variability from sources other than the channel, for example speaker and background variabilities.
	
	Figure~\ref{fig:num_enroll_segs} demonstrates the impact of the number of enrollment segments on speaker recognition performance. The top and bottom DET plots show the performance curves for a top performing system as a function of the number of enrollment segments (1 versus 3) form CMN2 and MLS trials, respectively. As expected, the larger the number of enrollment segments the better the performance. 
	


	Figure~\ref{fig:duration} shows DET plots for the various test segment speech durations (10--60~s) in the CTS Challenge, for the CMN2 (top) and MLS (bottom) trials. Results are shown for a top performing submission. Different performance trends are observed for the CMN2 versus MLS. For CMN2, we see a sharp drop in performance when the segment speech durations decrease from 20 seconds to 10 seconds, while for other durations, performance gaps remain small, in particular in the operating point of interest (i.e., the low false-alarm region). On the other hand, although we observe limited performance difference is for durations longer than 40~s for MLS, there is a rapid drop in performance when the speech duration decreases from 30~s to 20~s, and similarly from 20~s to 10~s.
	
	\section{Conclusion}
	NIST has been conducting the latest iteration of the CTS Challenge since August 2020. Similar to the SRE19 CTS Challenge, it is a leaderboard-style speaker recognition evaluation with \textit{Progress} and \textit{Test} subsets using CTS data extracted from unexposed telephony recordings from CMN2 and MLS corpora collected by the LDC. The \textit{Progress} leaderboard is live while the \textit{Test} leaderboard is updated periodically. In this paper, we presented a snapshot summary of the CTS Challenge (including the task, data, performance metric, the baseline system, as well as results and performance analyses) as of December 2021. The CTS Challenge has served, and will continue to serve, as a prerequisite for the regular SREs. Over the course of the Challenge, from August 2020 through December 2021, we observed remarkable improvements in speaker recognition performance for several top performers. The improvements are largely attributed to 1) the use of different variations of ReNet architectures along with different flavors of angular margin losses, 2) data augmentation, 3) data selection, and 4) long-duration fine-tuning.

	\section{Disclaimer}
	
	The results presented in this paper are not to be construed or represented as endorsements of any participant's system, methods, or commercial product, or as official findings on the part of NIST or the U.S. Government.
	
	Certain commercial equipment, instruments, software, or materials are identified in this paper in order to specify the experimental procedure adequately. Such identification is not intended to imply recommendation or endorsement by NIST, nor is it intended to imply that the equipment, instruments, software or materials are necessarily the best available for the purpose.

    \section{Acknowledgement}
	
	Experiments and analyses were performed, in part, on the NIST Enki HPC cluser.
	
	\balance
	\bibliographystyle{IEEEtran}
	\bibliography{Odyssey2020_BibEntries}
	
\end{document}